      [1999/12/01 v1.4c Il Nuovo Cimento]
\documentclass{cimento}

\newcommand{\ltsima} {$\; \buildrel < \over \sim \;$}
\newcommand{\gtsima} {$\; \buildrel > \over \sim \;$}
\newcommand{\lta} {\lower.5ex\hbox{\ltsima}}
\newcommand{\gta} {\lower.5ex\hbox{\gtsima}}

\newcommand{\be}{\begin{equation}}
\newcommand{\ee}{\end{equation}}
\newcommand{\ba}{\begin{eqnarray}}
\newcommand{\ea}{\end{eqnarray}}
\newcommand{\siml}{\lower4pt \hbox{$\buildrel < \over \sim$}}
\newcommand{\simg}{\lower4pt \hbox{$\buildrel > \over \sim$}}
\usepackage{graphicx}  
\title{Global properties of X-ray afterglows of GRB}
\author{L.~Piro\from{ins:x}} \instlist{\inst{ins:x}Istituto Astrofisica
Spaziale e Fisica Cosmica - Rome, INAF }
\PACSes{\PACSit{95.85.Nv}{X--rays} \PACSit{98.70.Rz}{gamma--ray
bursts}}
\begin{document}

\maketitle

\begin{abstract}
In this paper we review the general properties of X-ray
afterglows. We discuss in particular on the powerful diagnostics
provided by X-ray afterglows in constraining  the environment and
fireball in normal GRB, and the implications on the origin of dark
GRB and XRF. We also discuss on the observed properties of the
transition from the prompt to the afterglow phase, and present a
case study for a late X-ray outburst interpreted as the onset of
the afterglow stage.

\end{abstract}

\section{X-ray features}
The presence of X-ray features is an issue with important
implications on the origin of the progenitor and on the
cosmological use of GRB (see \cite{p04} for a review). So far,
different authors reported evidence, ranging from 2.8 to 4.7 sigma
significance, in six objects for features associated to iron
complex (see \cite{p04} and reference therein), and in 3-4 objects
for lines associated to lower Z elements (Mg, Si, S,
Ar)\cite{r+02,wro+02,bmr+03,wrh+03}. For what regards iron
features, when an independent redshift measurement is available,
the emission features in the afterglow phase are consistent with
highly ionized iron, while in the prompt phase the absorption
feature corresponds to neutral stage, suggesting a temporal
evolution guided by the large increase of ionization produced by
GRB photon field \cite{pl98}. The strong dependence of the line
efficiency on the ionization stage of the material, variable over
orders of magnitude from the prompt to the afterglow phase, can
explain the transitory presence of lines in a same object or the
tighter upper limits derived in some other bursts. In this
framework soft X-ray lines and iron lines are mutually exclusive,
because their maximum efficiencies are achieved at a different
ionization parameter \cite{lrr02}. While the present body of
observations can thus be consistent with theoretical expectations,
the former is still sparse, and more data, of higher statistical
quality, are required. Here we briefly discuss about the methods
to assess the statistical significance of these features. Usually
this is derived by using the F-test.
 According to \cite{pvc+02} the F-variable does not follow
the F distribution when the boundary of the normalization of an
additional model component is zero, as it is the case of emission
or absorption features. In such a case the correct probability
distribution of the F-variable is to be derived by using
MonteCarlo (MC) simulations of the null (i.e. the continuum)
model. For each MC realization of the null model, the value of the
F-variable is then derived by fitting the simulated spectrum with
the continuum model and then with the addition of the line
(hereafter we refer to this method with SP, standard
prescription). \cite{shr05} carried out a systematic analysis of
several afterglows, including all those with a reported evidence
of features. However they did not apply the aforementioned SP but
devised a different method, obtaining a much lower statistical
significance than reported in "discovery" papers. They claim that,
while the SP is in principle correct, in practice it fails in a
blind search of several lines, due to the inadequacy of $\chi^2$
minimization routines to find the absolute minimum. We have
applied the SP  to a single line at a given energy, i.e. a case
not affected by minimization convergence, specifically to the case
of GRB970508. The results of the different methods are thus
compared {\it under the same assumption}, i.e. single trial
probability.  Our result with the SP is presented in Fig.1. It
gives a statistical significance of 99.6\%. This is actually
slightly better  than the 99.3\%, originally derived in
\cite{pcf+99} by applying the F-test. On the contrary \cite{shr05}
derive a significance of 60\%. The same data were also reanalyzed
in \cite{pvc+02}, which derive a 99.3\% significance by applying
the SP, consistent with our estimation.

\begin{figure}
\includegraphics[width=7cm,origin=c,angle=0]{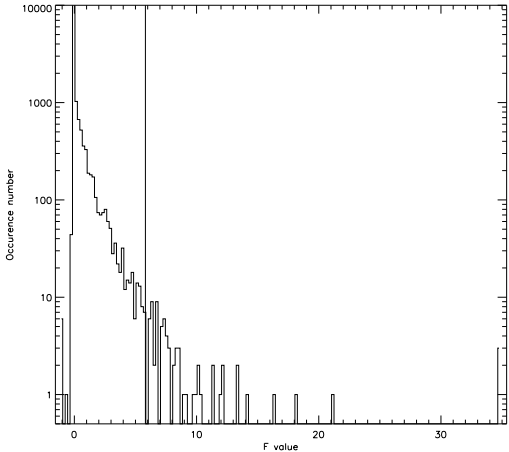}
\vspace{0.0cm}
 \caption{MonteCarlo simulations of F variable distribution for the
 BeppoSAX observation of GRB970508 for the null model (power law without
 line). The vertical line identifies the
 observed value of the F variable after the addition of a line,
 that corresponds to a probability of a chance fluctuation of 0.4\% }
 \label{fig_1}
\end{figure}

As originally noted in \cite{pvc+02} , the F-test slightly {\it
underestimates} the significance of an emission feature. We also
note that in XSPEC the $\chi^2$ computation with the default
setting adopts errors derived from observed counts, while the
correct formula (that can be selected in the program, but requires
the background file) requires the variance to be computed from the
model \cite{wdj+95}. This leads to an underestimation of the
significance of deviations above the continuum, i.e. of emission
lines, and to an overestimation of deviations below the continuum
(absorption feature) the more severe when the equivalent width is
large and the spectrum is source-dominated. We therefore conclude
that the statistical significance derived by using the F-test, at
least in the case of single line searched in a narrow range, gives
a conservative estimation of the confidence level of an emission
feature. An extension of the work  to blind searches for multiple
lines (soft X-ray lines) is in progress.

\begin{figure}
\vspace{2.0 cm}
\includegraphics[height=10cm,width=10cm,origin=c,angle=0]{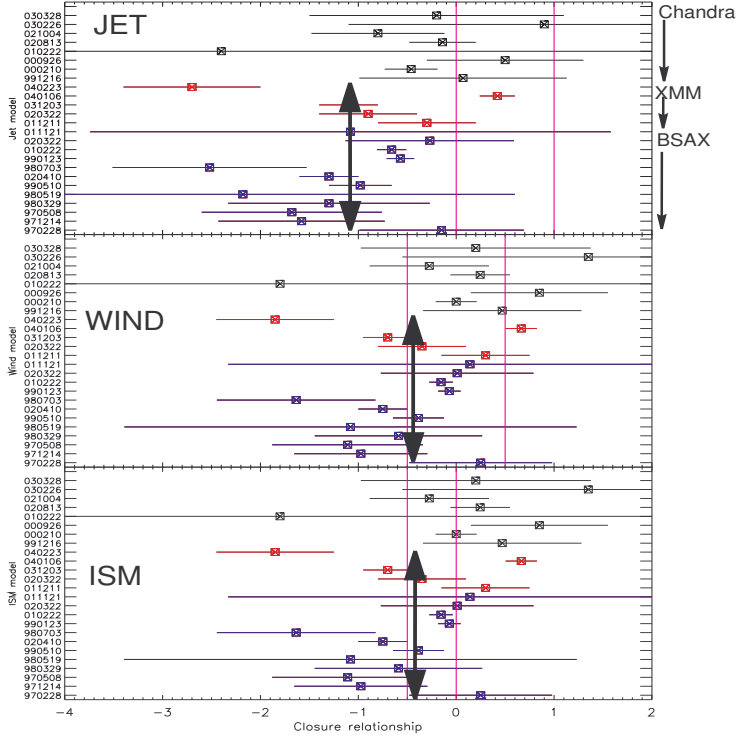}
\vspace{0.0cm}
 \caption{Distribution of the values of the closure relationships for jet (upper panel), spherical
 expansion in ISM (mid panel)  and wind (lower panel).  The sample includes
 afterglows observed with Chandra, XMM and BeppoSAX. The vertical lines
 are the expected values for $\nu > \nu_c$ (left) and $\nu < \nu_c$ (right).
 The arrows identify the
 average value derived for the combined BeppoSAX and XMM sample.
 They are consistent
 with ISM or wind expansion with $\nu > \nu_c$ }
 \label{fig_2}
\end{figure}

\begin{figure}
\vspace{2.0 cm}
\includegraphics[width=7cm,origin=c,angle=0]{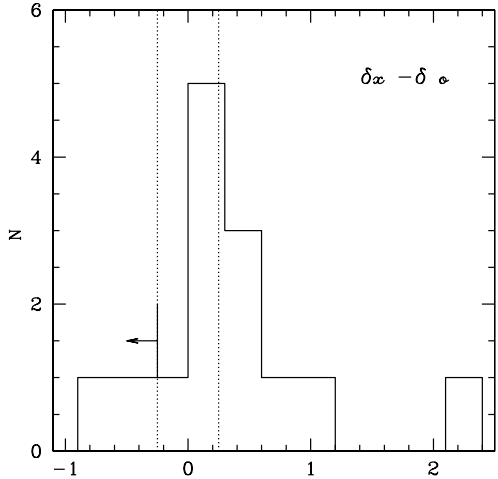}
\vspace{0.0cm}
 \caption{Distribution of the difference of decay indeces of X-ray and optical
 afterglows ($\delta_X-\delta_O$) for the BeppoSAX sample. The two dashed lines
 identify the expected value for a ISM ($\delta_X-\delta_O$=0.25)
and a wind ($\delta_X-\delta_O$=-0.25)}
 \label{fig_3}
\end{figure}

\section{Afterglow evolution and constraints on the fireball and
environment}

In previous papers \cite{p04,dpp+03}we have shown that the
application of closure relationship derived from spectral and
temporal evolution of X-ray afterglows by BeppoSAX set relevant
contraints on the fireball model and the environment. In
particular we derived that the fireball at $t\leq$ 1-2 day does
not (yet) show evidence of collimated flow, and is consistent with
a spherical expansion with a cooling frequency below the X-ray
range. We are extending this analysis including the XMM and
Chandra observations. The results are summarized in Fig.2. We note
that the XMM data are fully consistent with the BSAX sample.
Indeed, the typical observing time by XMM is similar to that of
BeppoSAX (few hours -- 2 days). In the case of Chandra, there is a
significant difference in the closure relationship, with the
Chandra sample consistent with a jet flow \cite{g+05}. This
apparent inconsistency is likely due to the fact that Chandra
observations start on average at later times, and that the effects
of jet flow on the light curve (i.e. the break time) take place
around 2-3 days.

Density profiles derived from broad-band afterglow modelling are
particularly intriguing, in that the majority of events are
consistent with a constant density environment, and only in few
cases a wind profile is clearly preferred \cite{cl00,pk02}. This
is at odd with the simple expectation of massive star progenitors.
Recently \cite{clf04} proposed a solution to solve this
discrepancy, arguing that a region of constant density would be
produced at the boundary of the wind with the molecular cloud
surrounding the progenitor.

What can we tell about this issue from X-ray afterglows?   When
the cooling frequency is above the X-ray band, as it happens in
the majority of the cases, the closure relationship for wind and
ISM are degenerate (Fig. 2). One notable exception is GRB040106
\cite{gpp04} where $\nu_c$ is below the X-ray band and the
combined spectral and temporal evolution in X-rays indicates a
wind profile. The addition of the optical temporal evolution
provides a powerful indicator to tell the ISM vs wind environment.
In a wind profile the decay in X-rays should be shallower than in
the optical, while the reverse holds true in a constant density
environment (ISM). The absolute value of the difference of the
decay slopes is 0.25. In Fig. 3 we plot the difference in decay
slopes $\delta_X-\delta_O$ for the BeppoSAX sample. This shows
that the ISM profile is preferred in most of the cases. One
intriguing outlier (see next section) is GRB011121 \cite{pds+05}.
In this event we derived $\delta_X=1.29 \pm 0.04$ vs
$\delta_O=1.66 \pm 0.06$ observed by \cite{pbr+02}, consistent
only with a wind profile, as also suggested in \cite{pbr+02} by
comparing optical and radio data. Another wind-profile candidate
is presented in \cite{gp05}.

\section{Early and delayed afterglows and the transition from the prompt phase}

Early papers combining the BeppoSAX WFC and NFI \cite{fac+00} have
outlined a clear separation in between two phases of the GRB
emission. The prompt phase is characterized by a hard spectrum
with a strong hard-to-soft spectral evolution. This is followed,
on a time scale of tens of seconds, by a second phase with a soft
spectrum, well described by a power law with spectral index
$\alpha\sim 1$ with no substantial spectral variation. This phase
is associated with the onset of the afterglow on the bases of two
evidences. First,  the X-ray spectrum is similar to that observed
in the late afterglow at 1 day. Second, the X-ray flux falls on
the backward extrapolation of the late afterglow light curve. The
prompt and early afterglow phases can be separated by temporal
gaps or can be mixed \cite{sdp+04} but, so far, the transition has
been always observed within tens of seconds from the onset of the
prompt emission.

In few cases, however, the transition appears to take place on a
much longer time scale. In a recent paper \cite{pds+05}
have presented evidence of an X-ray burst starting hundreds of
seconds after the prompt phase in two bursts, GRB011121 and
GRB011211 (see also \cite{gp05} for XRF011030). In the November
burst the spectrum of the late X-ray burst was markedly softer
than that of the preceding emission and  was similar to that
observed in the late afterglow observations. It is therefore
tempting to identify the late X-ray bursting as the onset of the
afterglow.  Contrary to what  observed  for transitions   on
shorter time scales, the decay part of the late X-ray burst cannot
be connected to the 1-day afterglow emission with a single
power-law $(t-t_0)^{-\delta_{\rm X}}$ (Fig. 4). {\it However, this
is the case (Fig. 4 right panel) when $t_0$ is set equal to the
onset of the late X-ray burst}. This empirical result can be
explained in the framework of the fireball model, taking into
account the thickness of the shell. The onset of external shocks
depends on the dynamical conditions of the fireball and, in
particular, two regimes can be identified depending on the
``thickness'' of the fireball \cite{sp99a}. In the regime of thin
shell, the reverse shock crosses the shell before the onset of the
self-similar solution (i.e. when an ISM mass $m=M_0/\Gamma_0$ is
collected, $M_0$ and $\Gamma_0$ being the rest mass and asymptotic
Lorentz factor of the fireball respectively). As a consequence,
the onset of the afterglow coincides with the deceleration time.
 Moreover, the evolution of the afterglow
after the peak is well described by a power-law decay, if the time
is measured starting from the explosion time, very well
approximated by the time at which the first prompt phase photons
are collected. In the case of a thick shell, the reverse shock has
not crossed the shell when the critical mass $m=M_0/\Gamma_0$ has
been collected, and therefore the external shock keeps being
energized for a longer time. The peak of the afterglow emission
therefore coincides  with the shell crossing time of the reverse
shock, equal to the duration of prompt phase.  The afterglow decay
will be well described with a single power-law only if the time is
measured starting from the time at which the inner engine turns
off, roughly coincident with the GRB duration.

\begin{figure}
\includegraphics[height=6.5cm,origin=c,angle=-90]{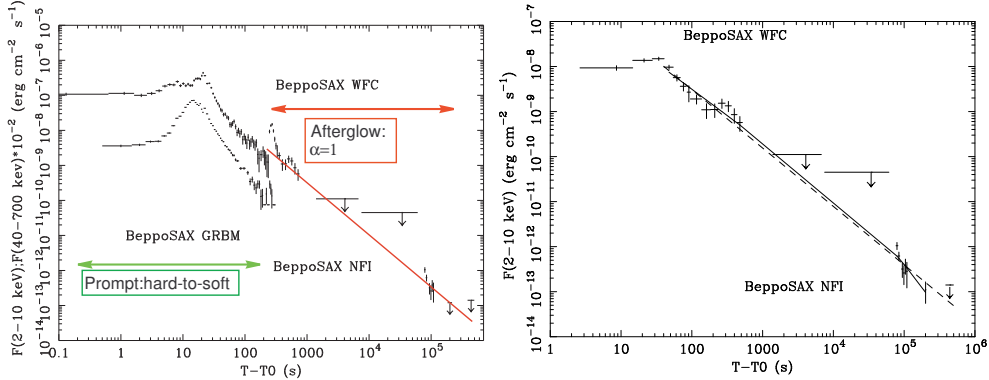}
\includegraphics[height=6.5cm,origin=c,angle=-90]{piro1_f5.eps}
\vspace{0.0cm}
 \caption{Light curves of GRB011121. On the left we show the curves
of WFC, NFI and GRBM when $t_0$ is, as usual, set with the start
of the prompt emission. Note that the outburst observed around 250
s lies above the power law describing the late afterglow emission.
However, when $t_0$ is shifted to the onset of the outburst (right
panel), the decay part of the outburst and the late afterglow fall
on a same power law
 }
 \label{fig_4}
\end{figure}

\section{Dark bursts}
One of the most discussed issues of the field is the origin of the
dark GRB \cite{dpp+03}. A first estimation of the fraction of
these events, around 50\% of the whole population of GRB, was
based adopting an upper limit on the optical magnitude at 1 day of
around 22-23. The availability of fast and precise position, as
e.g. delivered by HETE2, has led to the detection of a few rapidly
fainting optical transients, that would likely be classified as
dark GRB with observations performed at later times \cite{fps+03}.
 Fynbo et al. \cite{fjg+01} showed that about 75\% of dark GRB were
consistent with no detection if they were similar to dim burst
detected in the optical with deep searches. Thus, the original
definition of a dark burst is affected by significant
observational bias.

A more effective classification makes use of broad band
information. Let us consider a first set of causes that can make a
burst optically dim. A fast decay, as it would be the case of a
highly collimated jet is one possibility.  They can also be
intrinsically under-luminous events, or GRB located at distances
higher than that of OTGRB, but at a redshift not greater than 5
(see below). {\it In all these cases the afterglow flux should
scale of the same factor at all wavelengths}. Indeed \cite{dpp+03}
found that dark GRB are on average 6 times fainter in X-rays that
OTGRB. Incidentally we note that this effect can account, at least
in part, for the higher number of OTGRB identified by HETE2.

 {\it On the contrary, the optical flux of a GRB at z$\gta$5
or of a GRB in a dusty star forming region should be depleted not
only in absolute magnitude but also with respect to other
wavelengths}. Guided by this consideration, \cite{dpp+03} have
carried out a study of dark GRB vs OTGRB comparing their X-ray vs
optical fluxes. In 75\% of dark GRB's, the upper limits on the
optical-to-X-ray flux ratio ($f_{OX}$) are consistent with the
ratio observed in OTGRB, which is narrowly distributed around a
optical-to-X-ray spectral index  $\beta_{OX}=0.8$ (that is, modulo
a constant factor, the same as $log(f_{OX})$). This population of
events is therefore consistent with being OTGRB going undetected
in the optical because searches were not fast or deep enough.
 However, for about 25\% of dark GRB, $f_{OX}$ is at least a
factor 5-10 lower than the average value observed in OTGRB, and
also lower than the smallest observed $f_{OX}$. In terms of
spectral index, these events have $\beta_{OX} < 0.6$. Furthermore,
the optical upper limits are also lower than the faintest optical
afterglow. These GRB cannot be therefore explained as dim OTGRB's,
and are named {\it truly dark or optically depleted GRB}.

We stress that the upper limit on $f_{OX}$ for optically depleted
GRB is model-independent, being derived by a comparison with the
optically bright GRB, where the $f_{OX}$ distribution is rather
narrow,  clustering around the average value within a factor of 2
(the 1 sigma width). A similar value on the upper limit on
$f_{OX}$ has been derived in two dark GRB (\cite{dfk+01,pfg+02})
by modelling the broad band data via the standard fireball model.
Both of these events have been associated with host galaxies at
z$\lta$5, leading to the conclusion that the optical is depleted
by dust in star-forming region.

A similar approach has been followed by \cite{jhf+04}. They
derived $\beta_{OX}$ for a large sample of GRB and compared it
with the expectations of the fireball model. They find that at
least 10\% of the objects of their sample have an optical flux (or
upper limits) fainter than the minimum allowed by the model,
corresponding to $\beta_{OX}< 0.5-0.55$, and similar to the {\it
observed} limit derived by \cite{dpp+03}. In conclusion,  $\approx
10-20 \%$ of the burst population is characterized by an optical
afterglow emission substantially fainter than that expected from
the X-ray afterglow flux. As mentioned above this behaviour cannot
be accounted by {\it achromatic} effects, such as jet expansion or
luminosity. It requires causes that are selectively depleting the
optical range with respect to X-rays, such as dust extinction in
star forming regions or absorption by Ly$_\alpha$ forest of GRB at
z$>$5. As noted above, in few events the likely explanation is
dust extinction. The detection of high-z GRB is more challenging,
since the light of the host galaxy should be extremely dim.  The
large number of SWIFT localizations should hopefully lead to the
first identifications of high-z GRB.

\section{X-ray flashes}

This new class of GRB was originally discovered by BeppoSAX
\cite{hzkw01},  confirmed and extended by HETE2 \cite{l05}. Their
origin is still to be understood and, on this issue,  we would like
to show some implications derived from the properties of  X-ray
afterglows of XRF when compared to those of normal GRB.  \cite{dp05}
find that the distribution of X-ray fluxes
(at 12 hours after the burst) of the two classes are consistent
with each other. In particular the ratio of the average flux of
the two populations is $1.2\pm0.6$. This result appears at odd
with simple expectations by two popular scenarios. Let us first
 assume that XRF are normal GRB lying at
larger distances. In such a case one would expect to observe a
fainter afterglow, assuming that no selection effect biases the
sample. For example, assuming  an average redshift of 5 for XRF vs
$z=1$  for GRB, the X-ray afterglow flux should be on average 7
times fainter in XRF, contrary to what is observed. Indeed we
already know that some XRF are much closer (see references in
\cite{dp05}). Nonetheless, about 50\% of the XRF for which optical
observations were carried out,  lack an optical counterpart and we
cannot exclude that some of these events are at $z\gta5$. A second
scenario is that XRF are normal GRB seen off-axis (e.g. \cite{l05}
for a review). Let us assume that the only difference between the
two classes is the viewing angle. This is equivalent to the
unification scenario of Seyfert galaxies in the strong form.
Again, under this assumption, the flux of the X-ray (and optical)
afterglow observed at 12 hours should be substantially fainter
than observed. In particular, both for the homogenous and for the
universal jet models \cite{dp05} find that the observed value of
the afterglow ratio requires  a maximum angle of a few degrees. In
conclusion, the average property X-ray afterglows of XRF appears
too bright to be consistent with a single origin, either in terms
of off-axis jet or high-z scenario. One possibility is that we are
missing a significant fraction of faint XRF and it is hoped that
this population could be probed with more data by HETE2 and SWIFT.
In order to assess alternative scenarios for the prompt emission
(\cite{mdb+04}) predictions on the afterglow properties need to be
made.



\acknowledgments I would like to thank B. Gendre, M. De Pasquale,
A. Corsi, A. Galli and V. D'Alessio for inputs on this paper.



\end{document}